\documentclass[aps,prd,twocolumn,superscriptaddress]{revtex4-1}

\usepackage[breaklinks=true]{hyperref}
\usepackage{graphicx}
\usepackage{amsmath}

\newcommand{\unit}[1]{~\mathrm{#1}}
\newcommand{\shiki}[1]{Eq.~(\ref{#1})}
\newcommand{\zu}[1]{Fig.~{\ref{#1}}}
\newcommand{\hyou}[1]{Table~{\ref{#1}}}
\newcommand{\sho}[1]{Sec.~{\ref{#1}}}

\newcommand{\kB}{k_{\rm B}}
\newcommand{\obj}{\mathcal{F}}
\newcommand{\attn}{{\rm attn}}
\newcommand{\SRM}{{\rm SRM}}
\newcommand{\ITM}{{\rm ITM}}
\newcommand{\ETM}{{\rm ETM}}
\newcommand{\arm}{{\rm arm}}
\newcommand{\f}{{\rm f}}
\newcommand{\m}{{\rm m}}
\newcommand{\s}{{\rm s}}
\newcommand{\coa}{{\rm c}}
\newcommand{\Ta}{{\rm Ta}}
\newcommand{\Si}{{\rm Si}}
\newcommand{\IM}{{\rm IM}}
\newcommand{\chirpmass}{\mathcal{M}_{\rm c}}
\newcommand{\n}{{\rm n}}
\newcommand{\rtHz}{{\rm \sqrt{Hz}}}
\newcommand{\Np}{N_{\rm p}}
\newcommand{\Nd}{N_{\rm d}}
\newcommand{\phidet}{\phi_{\det}}
\newcommand{\Msun}{M_{\odot}}

\begin{document}

\title{Particle swarm optimization of the sensitivity of a cryogenic\\ gravitational wave detector}

\author{Yuta~Michimura}
  \email{michimura@granite.phys.s.u-tokyo.ac.jp}
  \affiliation{Department of Physics, University of Tokyo, Bunkyo, Tokyo 113-0033, Japan}
\author{Kentaro~Komori}
  \affiliation{Department of Physics, University of Tokyo, Bunkyo, Tokyo 113-0033, Japan}
\author{Atsushi~Nishizawa}
  \affiliation{Kobayashi-Maskawa Institute for the Origin of Particles and the Universe, Nagoya University, Nagoya, Aichi 464-8602, Japan}
\author{Hiroki~Takeda}
  \affiliation{Department of Physics, University of Tokyo, Bunkyo, Tokyo 113-0033, Japan}
\author{Koji~Nagano}
  \affiliation{Institute for Cosmic Ray Research, University of Tokyo, Kashiwa, Chiba, 277-8582, Japan}
\author{Yutaro~Enomoto}
  \affiliation{Department of Physics, University of Tokyo, Bunkyo, Tokyo 113-0033, Japan}
\author{Kazuhiro~Hayama}
  \affiliation{Department of Applied Physics, Fukuoka University, Nanakuma, Fukuoka 814-0180, Japan}
\author{Kentaro~Somiya}
  \affiliation{Department of Physics, Tokyo Institute of Technology, Meguro, Tokyo 152-8550, Japan}
\author{Masaki~Ando}
  \affiliation{Department of Physics, University of Tokyo, Bunkyo, Tokyo 113-0033, Japan}
  \affiliation{National Astronomical Observatory of Japan, Mitaka, Tokyo, 181-8588, Japan}
  \affiliation{Research Center for the Early Universe, University of Tokyo, Bunkyo, Tokyo 113-0033, Japan}
\date{\today}

\begin{abstract}
Cryogenic cooling of the test masses of interferometric gravitational wave detectors is a promising way to reduce thermal noise. However, cryogenic cooling limits the incident power to the test masses, which limits the freedom of shaping the quantum noise. Cryogenic cooling also requires short and thick suspension fibers to extract heat, which could result in the worsening of thermal noise. Therefore, careful tuning of multiple parameters is necessary in designing the sensitivity of cryogenic gravitational wave detectors. Here, we propose the use of particle swarm optimization to optimize the parameters of these detectors. We apply it for designing the sensitivity of the KAGRA detector, and show that binary neutron star inspiral range can be improved by 10\%, just by retuning seven parameters of existing components. We also show that the sky localization of GW170817-like binaries can be further improved by a factor of 1.6 averaged across the sky. Our results show that particle swarm optimization is useful for designing future gravitational wave detectors with higher dimensionality in the parameter space.
\end{abstract}


\maketitle

\section{Introduction}
The first direct detections of gravitational waves from binary black holes~\cite{GW150914} and binary neutron star systems~\cite{GW170817,Multimessenger} by Advanced LIGO~\cite{aLIGO} and Advanced Virgo~\cite{AdV} have opened a vast new frontier in physics and astronomy. Improving the sensitivity of these interferometric detectors would increase the number of detections and enable better sky localization and more precise binary parameter estimation~\cite{ObservationScenario}. The designed sensitivity of state of the art gravitational wave detectors is limited by seismic noise, thermal noise and quantum noise, and there have been extensive studies to reduce these fundamental noises~\cite{RanaRMP,NewtonianNoise,GrasCoating,HammondSuspension,ShapiroCryogenics,Kimble2001,LIGOSqueezing,FilterCavity,SomiyaParametric,InternalSqueezing,EPR,LIGO-LF,LIGO-HF}.

For thermal noise reduction, KAGRA~\cite{SomiyaKAGRA,AsoKAGRA,iKAGRA,MichimuraKAGRA} and some proposals of future gravitational wave detectors~\cite{ET,SheilaLong,CE} plan to cool the test mass mirrors to cryogenic temperatures. Cryogenic cooling in gravitational wave detectors is not straightforward since incident laser power to the test masses is in the order of a megawatt to reduce quantum shot noise. The heat extraction is done by the fibers suspending the test mass. In terms of heat extraction efficiency, the fibers should be short and thick, but in terms of thermal noise, fibers should be long and thin to effectively dilute the mechanical loss of the pendulum~\cite{SaulsonSuspension,Cumming}.

Therefore, to design the sensitivity of cryogenic gravitational wave detectors, parameters related to thermal noise and those related to quantum noise must be carefully tuned simultaneously. The sensitivity design will be an optimization problem in highly multidimensional parameter space. Future gravitational wave detectors will have more parameters to be optimized when quantum noise reduction techniques such as squeezed vacuum injection~\cite{LIGOSqueezing}, filter cavity~\cite{FilterCavity}, parametric amplifier~\cite{SomiyaParametric}, and intra-cavity optomechanical filtering~\cite{LIGO-HF} are applied. In this situation, classical grid-based searches will be computationally expensive, and stochastic approaches must be explored.

Here, we demonstrate the use of particle swarm optimization (PSO)~\cite{PSO1995} in this context.  As PSO is a stochastic method, unlike grid-based search, the computational cost for searching the global maximum does not grow exponentially with the dimensionality of the parameter space. However, like other stochastic methods, convergence to the global maximum is guaranteed only in the limit of infinite sampling. Compared with other stochastic methods such as genetic algorithms, an attractive feature of PSO is that it has a small number of design variables. PSO can be designed by just determining the number of particles and termination criterion. The only prior information required is the search boundary in the parameter space.

Being a metaheuristic algorithm, PSO has been applied to wide range of areas including astronomy. Previous studies show that PSO is effective for astronomical applications such as orbital study of galactic potentials~\cite{PSOOrbit}, gravitational lens modeling~\cite{PSOLensing}, cosmological parameter estimation using cosmic microwave background data~\cite{PSOCMB}, and gravitational wave data analysis~\cite{PSOMohanty2010,PSOBouffanais2016,PSOMohanty2017}. In this paper, we show that PSO is also effective for the detector design by applying it for the sensitivity optimization of the KAGRA cryogenic gravitational wave detector.

The rest of this paper is organized as follows. In \sho{SecSensitivity}, we describe the KAGRA sensitivity calculation and define the detector parameters used for optimization. We then define the objective function to be maximized in \sho{SecFunction}. For the objective function, we studied two cases: binary neutron star inspiral range and sky localization error of GW170817-like binary. The algorithm of PSO and our procedure for tuning the design variables of PSO is discussed in \sho{SecPSO}. Section~\ref{SecResults} presents our results of the sensitivity optimization. Our conclusions and prospects are summarized in \sho{SecConclusion}.

\section{KAGRA sensitivity calculation} \label{SecSensitivity}
\begin{figure}
\begin{center}
\includegraphics[width=\hsize]{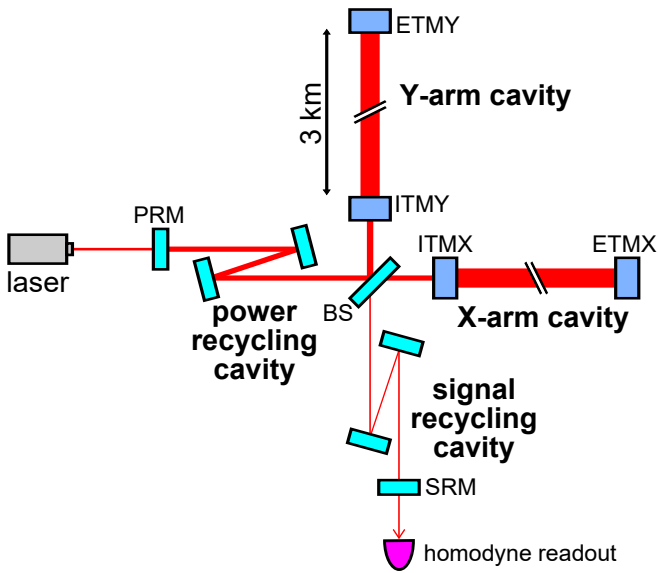}
\end{center}
\caption{\label{KAGRAInterferometer} Schematic of the KAGRA interferometer. ITMs and ETMs are cryogenic sapphire mirrors, while all the other mirrors are fused silica mirrors at room temperature. The gravitational wave signal is extracted from the photodiode detecting the transmitted light of SRM.}
\end{figure}

KAGRA is an interferometric gravitational wave detector located at an underground site in Japan. As shown in \zu{KAGRAInterferometer}, the KAGRA interferometer is a resonant sideband extraction interferometer~\cite{MizunoRSE} similar to Advanced LIGO and Advanced Virgo. Advanced LIGO and Advanced Virgo are room temperature interferometers, but KAGRA has two 3-km long arm cavities formed by cryogenic sapphire test masses. The beam splitter (BS) and two arm cavities form a Michelson interferometer, which is sensitive to the differential arm length change caused by gravitational waves. A power recycling mirror (PRM) is added to effectively increase the input power by 10. A signal recycling mirror (SRM) is added to form a signal recycling cavity (SRC) with main arm cavities to shape the quantum noise by tuning the SRM reflectivity and SRC length~\cite{BuonannoChen2001}.

The main sapphire mirrors, input test masses (ITMs) and end test masses (ETMs) are suspended by a 14-m long eight-stage pendulum to attenuate the displacement noise from ground motion~\cite{HiroseTypeA,Actuator}. The last four stages of the system are cooled down to cryogenic temperatures at around 20~K~\cite{Cryopayload} and are critical for the sensitivity design (see \zu{KAGRACryopayload}).

In this section, we describe the details of the KAGRA sensitivity calculation by describing seismic noise, mirror thermal noise, suspension thermal noise, and quantum noise. The sensitivity spectrum over Fourier angular frequency $\omega=2 \pi f$ can be calculated as
\begin{equation}
 S_{\n}(\omega) = S_{\rm seis}(\omega) + S_{\rm mir}(\omega) + S_{\rm susp}(\omega) + S_{\rm quant}(\omega).
\end{equation}
Throughout this paper, $S(\omega)$ denote one-sided power spectral density in strain (1/Hz).

We also discuss the heat extraction through suspension fibers and summarize the parameters to optimize. The calculation basically follows the work done in Ref.~\cite{SomiyaKAGRA}, but is updated to incorporate the design change in the cryogenic mirror suspension system~\cite{Cryopayload}. The fixed detector parameters and parameters used for our optimization are summarized in \hyou{FixedParams} and \hyou{SearchRange}, respectively.

\subsection{Seismic noise}
The mirror displacement noise due to the ground motion attenuated through the test mass suspension system can be approximated as~\cite{SomiyaKAGRA}
\begin{align}
 S_{\rm gnd}(\omega) & = (1.6 \times 10^{-16} \unit{/\rtHz})^2 \times  \nonumber \\
 & \left[ \left(\frac{0.58\unit{Hz}}{f} \right)^{11.4} + \left(\frac{1\unit{Hz}}{f} \right)^{13.6} + \left(\frac{1.2\unit{Hz}}{f} \right)^{16} \right] .
\end{align}
Ground motion also causes fluctuation of the gravity field, which disturbs the mirror motion. This is called Newtonian noise~\cite{NewtonianNoise}, and the simulated Newtonian noise from the surface and bulk motion of the mountain containing KAGRA is approximated as~\cite{SomiyaKAGRA}
\begin{equation}
 S_{\rm NN}(\omega) = (4 \times 10^{-20} \unit{/\rtHz})^2 \times \left(\frac{1 \unit{Hz}}{f} \right)^8 .
\end{equation}

The total seismic noise will be
\begin{equation}
  S_{\rm seis}(\omega) = S_{\rm gnd}(\omega) + S_{\rm NN}(\omega).
\end{equation}
In reality, seismic noise slightly changes if test mass suspension fiber parameters are changed. However, seismic noise is more than an order of magnitude lower than other noises in the observation band above 10~Hz, and this effect is negligible. We therefore fixed the seismic noise level for our optimization process.

\subsection{Mirror thermal noise}
The Brownian motion of the test mass surface from mechanical losses is a limiting noise source in the mid-frequencies of the observation band. The mirror substrate Brownian noise~\cite{LevinSubstrate} and coating Brownian noise~\cite{HarryCoating} can be calculated by
\begin{equation}
  S_{\rm sub}(\omega) = \frac{4 \kB T_\m}{\omega L_\arm^2} \frac{\phi_\m}{\sqrt{\pi} w} \frac{1-\sigma_\m^2}{Y_\m}
\end{equation}
and
\begin{align}
  S_{\rm coa}(\omega) &= \sum_{c=\Si,\Ta} \frac{4 \kB T_\m}{\omega L_\arm^2} \frac{d_\coa \phi_\coa}{\pi w^2} \times \nonumber \\
& \frac{Y_\coa^2(1+\sigma_\m^2)^2(1-2\sigma_\m)^2+Y_\m^2(1+\sigma_\coa)^2(1-2\sigma_\coa)}{Y_m^2 Y_c (1-\sigma_\coa^2)} ,
\end{align}
respectively. Here, $\kB$, $\sigma$ and $Y$ are the Boltzmann constant, Poisson ratio and Young's modulus, respectively, with the subscript indicating mirror substrate for $\m$ and coating for $\coa$. KAGRA uses alternating silica/tantala coating~\cite{HiroseCoating,YamamotoCoating} and the total coating thermal noise is a sum of noises from silica layers ($\Si$) and tantala layers ($\Ta$).

Thermal expansion of the mirror substrate due to temperature fluctuation from diffusion losses cause thermoelastic noise. Thermoelastic noise at cryogenic temperatures is approximately given by~\cite{CerdonioThermoelastic,SomiyaThermoelastic}
\begin{equation}
 S_{\rm TE}(\omega) \simeq \frac{4 \kB T_\m^2 (1+\sigma_\m)^2 \alpha_\m^2}{L_\arm^2 \sqrt{\pi \kappa_\m C_\m \omega}},
\end{equation}
with $\alpha_\m$, $\kappa_\m$, and $C_\m$ being linear thermal expansion, thermal conductivity and specific heat per volume, respectively. To treat the temperature dependence of these three parameters, we used fitted functions of measured values reported in Refs.~\cite{ThermalExpansion,UchiyamaCLIO,ThermophysicalProperties}.

The total mirror thermal noise will therefore be the sum of all the noises above for all four test masses:
\begin{equation}
 S_{\rm mir}(\omega) = 2 \sum_{\rm{ITM,ETM}} \left( S_{\rm sub}(\omega) + S_{\rm coa}(\omega) + S_{\rm TE}(\omega) \right).
\end{equation}
Coating thermo-optic noise is low at cryogenic temperatures and is thus ignored here~\cite{YamamotoThermoOptic}.

\begin{table}
\caption{\label{FixedParams} Fixed KAGRA detector parameters used for the sensitivity calculation. Parameters without a reference come from either Ref.~\cite{SomiyaKAGRA} or actual measurement.}
\begin{ruledtabular}
\begin{tabular}{ll}
 & Value  \\
\hline
arm length & $L_{\arm}=3000 \unit{m}$ \\
ITM transmittance & $T_{\ITM}= 0.4\%$ \\
laser wavelength & $\lambda= 1064 \unit{nm}$ \\
\multicolumn{2}{l}{Sapphire test mass} \\
\quad radius & $r_{\m}=11 \unit{cm}$ \\
\quad thickness & $t_{\m}=15 \unit{cm}$ \\
\quad mass & $m_{\m} = 22.8 \unit{kg}$ \\
\quad loss angle~\cite{UchiyamaLoss} & $\phi_{\m}=1.0 \times 10^{-8}$ \\
\quad absorption & $\beta_{\m}=50 \unit{ppm/cm}$ \\
\multicolumn{2}{l}{Silica/tantala coating} \\
\quad beam radius & $w=3.5 \unit{cm}$ \\
\quad thickness for ITM & $d_{\Si,\Ta}^{\ITM} =2.21, 1.44 \unit{\mu m} $ \\
\quad thickness for ETM & $d_{\Si,\Ta}^{\ETM} =3.87, 2.61 \unit{\mu m} $ \\
\quad loss angle~\cite{SomiyaKAGRA,HiroseCoating} & $\phi_{\Si,\Ta}=3.0 \times 10^{-4}, 5.0 \times 10^{-4}$ \\
\quad absorption & $\beta_{\coa}=0.5 \unit{ppm}$ \\
\multicolumn{2}{l}{Intermediate mass suspension (CuBe)} \\
\quad mass & $m_{\IM}=20.5 \unit{kg}$ \\
\quad temperature & $T_{\IM}=16 \unit{K}$ \\
\quad length & $l_{\IM}=26.1 \unit{cm}$ \\
\quad diameter & $d_{\IM}=0.6 \unit{mm}$ \\
\quad loss angle~\cite{NewmanG} & $\phi_{\IM}=5 \times 10^{-6}$ \\
\multicolumn{2}{l}{Sapphire blade spring} \\
\quad mass & $m_{\rm B}=55 \unit{g}$ \\
\quad temperature & $T_{\rm B}=T_{\IM}=16 \unit{K}$ \\
\quad loss angle~\cite{DanPhD} & $\phi_{\rm B}=7 \times 10^{-7}$ \\
\multicolumn{2}{l}{Test mass suspension (sapphire)} \\
\quad loss angle~\cite{DanPhD} & $\phi_{\IM}=2 \times 10^{-7}$ \\
\end{tabular}
\end{ruledtabular}
\end{table}

\subsection{Suspension thermal noise}
Contribution from the Brownian motion of the suspension system is significant at low frequencies. The power spectrum of the suspension thermal noise of a simple pendulum above its resonant frequency is approximated by~\cite{SaulsonSuspension}
\begin{equation} \label{SuspensionThermal}
 S_{\rm susp} (\omega) = \frac{4 \kB T_\f}{m \omega^5} \sqrt{\frac{4 \pi Y_\f g}{m}} \left( \frac{d_\f}{l_\f} \right)^2 \phi_\f,
\end{equation}
where $g$, $T_\f$, and $Y_\f$ is gravitational acceleration, temperature and Young's modulus of the suspension fiber, respectively. Since the suspension fiber of cryogenic test mass is tasked with heat extraction, $T_\f$ is not uniform across the fiber. However, it is shown by Ref.~\cite{KomoriSuspension} that it is safe to use the averaged temperature of the top ($T_\IM$) and the bottom ($T_\m$) ends of the fiber such that
\begin{equation}
  T_\f = \frac{T_\m + T_{\IM}}{2},
\end{equation}
because the elastic energy is distributed symmetrically, and is mostly stored at the both ends of the fiber.

\begin{figure}
\begin{center}
\includegraphics[width=\hsize]{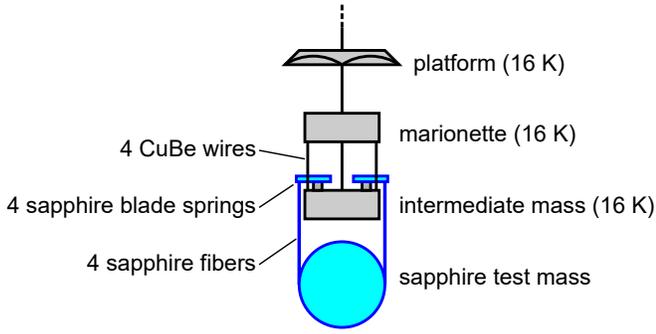}
\end{center}
\caption{\label{KAGRACryopayload} Schematic of the cryogenic test mass suspension system. The platform is suspended from upper room temperature vibration isolation stages.}
\end{figure}

Figure~\ref{KAGRACryopayload} shows the cryogenic stages of the KAGRA test mass suspension system. The sapphire test mass is suspended by four sapphire fibers from four sapphire blade springs attached to the intermediate mass. The intermediate mass is in turn suspended by four CuBe wires from the marionette. The marionette is then suspended by one maraging steel wire from the platform, which is suspended from upper room temperature stages. The intermediate mass, the marionette, and the platform are attached with high purity aluminum heat links from cryocoolers and are cooled down at $16\unit{K}$~\cite{MichimuraKAGRA,SakakibaraCryo}.

For the actual suspension thermal noise calculation, we used the modified version of the model developed for Virgo suspensions~\cite{PPPnote}. The model treats the triple pendulum consisting of the intermediate mass, the blade spring, and the test mass, and all the mechanical losses from their suspension wires and blade springs are included. It also treats coupling from the vertical thermal noise, which mainly comes from the blade spring and the intermediate mass suspension. The detailed calculation of the KAGRA suspension thermal noise is described in Ref.~\cite{KomoriSuspension}.

\subsection{Quantum noise}
The quantum noise comes from quantum fluctuation of light, and is a fundamental limit of interferometric gravitational wave detectors. The quantum noise of the detector without SRM is given by~\cite{Kimble2001}
\begin{equation} \label{QuantumNoise}
  S_{\rm quant}(\omega) = \frac{4 \hbar}{m \omega^2 L_\arm^2} \left( \frac{1}{\mathcal{K}} + \mathcal{K} \right),
\end{equation}
where
\begin{equation}
 \mathcal{K} = \frac{16 \pi c I_0}{m \lambda L_\arm^2 \omega^2 (\gamma^2+\omega^2)} .
\end{equation}
Here, $c$, $\hbar$, $I_0$ and $\gamma$ are the speed of light, Dirac's constant, input power to the BS, and arm cavity pole, respectively. Since most of the optical losses of the arm cavity comes from the transmission of ITM, $\gamma$ is given by
\begin{equation}
 \gamma = \frac{c T_{\ITM}}{4 L_\arm} .
\end{equation}

The first term is called shot noise and it comes from the quantum fluctuation of laser power incident on the detection photodiode. The second term in \shiki{QuantumNoise} is called radiation pressure noise, which comes from the mirror displacement caused by the quantum fluctuation of laser power incident on the mirror. There is a trade-off between radiation pressure noise and shot noise since the former is proportional to, and the latter is inversely proportional to the input power. By tuning the readout quadrature by homodyne detection, cancellation of these two noises is possible~\cite{SomiyaKAGRA}. Also, the addition of the SRC and its detuning make it possible to effectively make the input power frequency-dependent so that we can tune the detector bandwidth.

For the actual noise calculation, we used Eq.~(5.13) in Ref.~\cite{BuonannoChen2001}, which includes not only the effect of SRM but also power losses in the interferometer. We assumed the round-trip loss in the arm cavity, the loss at SRM, and the loss at the photodiode to be 100~ppm, 2000~ppm, and 10\%, respectively.

\subsection{Heat extraction and input power}
In cryogenic interferometers, quantum noise cannot be calculated independently from suspension parameters because the maximum input power allowed is dependent on the heat extraction capability of the fibers. The extractable heat of the fibers is given by
\begin{equation}
K_\f = \frac{N_\f \pi d_\f^2}{4 l_\f} \int_{T_\IM}^{T_\m} \kappa_\f(d_\f,T) {\rm d} T,
\end{equation}
where $N_\f=4$ is the number of fibers suspending the test mass, and $\kappa_\f(d_\f, T)$ is the thermal conductivity of the fiber. We used the measured thermal conductivity of sapphire which can be approximated with~\cite{SapphireFiber}
\begin{equation}
 \kappa_\f (d_\f,T) = 5800 \unit{W/m/K} \times \left( \frac{d_\f}{1.6 \unit{mm}} \right) \left( \frac{T}{20 \unit{K}} \right)^{2.2} .
\end{equation}
The thermal conductivity of sapphire below $\sim 40 \unit{K}$ is limited by boundary scattering of phonons and is proportional to the fiber diameter $d_\f$~\cite{TomaruFiber}.

On the other hand, the heat absorbed by the test mass, especially the ITM, is
\begin{equation}
  K_{\rm abs} = 2 \beta_\m t_\m I_{\ITM} + \beta_{\coa} I_{\rm circ} + K_{\rm rad} ,
\end{equation}
where $I_{\ITM}=I_0/2$ is the incident power to the ITM, and $I_{\rm circ}=4 I_\ITM/T_\ITM$ is the circulating power inside the arm cavity. The first term is the heat absorbed by the substrate and the second term is the heat absorbed by the coating. $K_{\rm rad}$ is additional heat introduced through the radiation from the apertures, and is estimated to be 50~mW~\cite{SakakibaraCryo}. The heat absorbed by the ETM is less than that of the ITM because the power of the laser beam that goes through the substrate is less by two orders of magnitude.

By requiring $K_\f > K_{\rm abs}$, maximum laser power at the BS can be calculated as
\begin{equation}
  I_0^{\max} = \frac{K_\f - K_{\rm rad}}{\beta_\m t_\m + 2 \beta_\coa/T_{\ITM}} .
\end{equation}
We can see that larger $d_\f$ and smaller $l_\f$ is better for reducing $T_\m$ and increasing $I_0$. However, as shown in \shiki{SuspensionThermal}, it also has the negative effect of increasing thermal noise.

For optimization, we introduced a power attenuation factor $I_{\attn}$ to calculate the input power,
\begin{equation}
  I_0 = I_{\attn} I_0^{\max}.
\end{equation}

\subsection{Parameters to optimize and their search ranges}
As shown in \hyou{SearchRange}, we have selected 7 parameters related to suspension thermal noise and quantum noise for optimization. These parameters are relatively easy to retune, even at the later stage of the detector commissioning. In particular, the first two parameters, $\phidet$ and $\zeta$, can be tuned without any additional investment to the detector. $T_\m$ and $I_{\attn}$ can also be tuned freely if enough power from the laser source is available. The change of $R_{\SRM}$ requires the replacement of the SRM. The last two, $l_\f$ and $d_\f$, requires the replacement of the last stage of the test mass suspension.

The search ranges of these parameters are determined based on experimental feasibility. The upper bound for $\phidet$ is set to $3.5^{\circ}$ since a highly detuned configuration can increase control noise~\cite{AsoLSC}. Here, $\phidet=0^{\circ}$ means the SRC is tuned, and $\zeta=90^{\circ}$ means a conventional readout in phase quadrature. The lower bound for $d_\f$ is determined considering the tensile strength of the fiber, and set to $0.8\unit{mm}$ to keep the safety factor to at least 3. The range for $T_\m$ is set to $[20,~30] \unit{K}$ so that temperature-dependent parameters can be approximated well with fitted functions of measured values.

The default KAGRA values of these parameters are also summarized in \hyou{SearchRange}. The latter four parameters are determined by practical reasons, and the first three parameters are determined based on a grid-based search to optimize the parameters to maximize the binary neutron star inspiral range~\cite{Latest201708}. Therefore, optimization including the latter four parameters could give an improved inspiral range. In addition, optimization for different objective functions should give different sets of parameters.

To study the effect of each parameter on KAGRA's sensitivity, we have tested three cases, varying the number of search parameters $\Nd$ to use for optimization. In the $\Nd=3$ case, we used only $\{\phidet,~\zeta,~T_\m\}$ for optimization and the other four parameters are fixed to their designed values. Similarly, in the $\Nd=5$ case, we used only $\{\phidet,~\zeta,~T_\m,~I_{\attn},~R_{\SRM}\}$ for optimization. Lastly, in the $\Nd=7$ case, we used all seven parameters.

\begin{table}
\caption{\label{SearchRange} The list of KAGRA detector parameters used for optimization. Their search ranges and default values~\cite{Latest201708} are as shown.}
\begin{ruledtabular}
\begin{tabular}{llcc}
 & & Search range & Default  \\
\hline
detuning angle (deg)   & $\phidet$   & $[0,~3.5]$    & 3.5   \\
homodyne angle (deg)   & $\zeta$     & $[90,~180]$   & 135.1 \\
mirror temperature (K) & $T_\m$      & $[20,~30]$    & 22    \\
power attenuation      & $I_{\attn}$ & $[0.01,~1]$   & 1     \\
SRM reflectivity (\%)  & $R_{\SRM}$  & $[50,~100]$   & 84.6  \\
fiber length (cm)      & $l_\f$      & $[20,~100]$   & 35    \\
fiber diameter (mm)    & $d_\f$      & $[0.8,~2.5]$  & 1.6   \\
\end{tabular}
\end{ruledtabular}
\end{table}

\section{Objective functions} \label{SecFunction}
To evaluate the sensitivity of the gravitational wave detector, we need a function for the figure of merit, and this will be our objective function to be maximized. Historically, the most commonly used figure of merit is the binary neutron star inspiral range. For multimessenger observations, source parameter estimation from gravitational wave signal will play a critical role~\cite{Multimessenger}.

Here, we consider two objective functions, the binary neutron star inspiral range, and the sky localization error of a binary neutron star, with similar parameters to GW170817.

\subsection{Inspiral range}
Once we choose a threshold for signal to noise ratio $\rho_{\rm th}$, we can derive a maximum distance at which we can see a binary inspiral signal. This distance is called the inspiral range and can be computed using the detector sensitivity $S_{\n}(f)$ by~\cite{Creighton}
\begin{equation} \label{InspiralRange}
 \mathcal{R} = \frac{0.442}{\rho_{\rm th}} \left( \frac{5}{6} \right)^{1/2} \frac{c}{\pi^{2/3}} \left( \frac{G \chirpmass}{c^3}  \right) \left[ \int_{f_{\min}}^{f_{\max}} \frac{f^{-7/3}}{S_{\n}(f)} {\rm d} f \right]^{1/2},
\end{equation}
where $G$ is the gravitational constant and $\chirpmass$ is the chirp mass of the binary. Using the component masses $m_1$ and $m_2$, the chirp mass is given by
\begin{equation}
 \chirpmass = \frac{(m_1 m_2)^{3/5}}{(m_1+m_2)^{1/5}}.
\end{equation}
The factor $0.442$ in \shiki{InspiralRange} is the sky average constant to average out the angular dependence of signal to noise ratio~\cite{SkyAverage}.

Following convention, here we consider the inspiral range for $m_1=m_2=1.4\Msun$ binary neutron star system, with $\rho_{\rm th}=8$, as one of our objective functions. We set the lower frequency end of the signal to noise ratio integration to be $f_{\min} = 10 \unit{Hz}$, and the upper end to be the gravitational wave frequency at the innermost stable circular orbit of the Schwarzshild metric
\begin{equation} \label{fISCO}
 f_{\max} = \frac{c^3}{6^{3/2} \pi G M_{\rm tot}},
\end{equation}
with $M_{\rm tot}=m_1+m_2$ being the total mass.

\subsection{Sky localization error}
The source parameter estimation performance of gravitational wave measurements can be evaluated using a Fisher information matrix~\cite{FIM1992,FIM1994}. The Fisher information matrix $\Gamma$ can be calculated using the derivatives of the waveform $h(f)$ with respect to source parameters $\lambda^i$ as
\begin{equation}
 \Gamma_{ij} = 4 \Re  \int_{f_{\min}}^{f_{\max}} \sum_k \frac{\partial h_k^{*}(f)}{\partial \lambda^i}  \frac{\partial h_k(f)}{\partial \lambda^j} \frac{{\rm d} f}{S_{\n,k}(f)} ,
\end{equation}
with
\begin{equation}
 h_k(f) = \mathcal{G}_k h (f).
\end{equation}
Here, $^{*}$ stands for complex conjugate, and $S_{\n,k}(f)$, $\mathcal{G}_k$, and $h_k(f)$ are sensitivity, geometrical factor, and waveform detected by the $k$-th detector, respectively. The geometrical factor accounts for the angular dependence of the signal, and is defined by
\begin{align}
 \mathcal{G}_k &= \frac{1}{2} \left[ (1+\cos^2{\iota}) F_{+,k}(\theta_\s,\phi_s,\psi_p) \right. \nonumber \\
 &\qquad \left. + 2i \cos{\iota} F_{\times,k} (\theta_\s,\phi_s,\psi_p) \right] e^{-i \phi_{D,k}(\theta_\s,\phi_s)} ,
\end{align}
where $\phi_{D,k}$ is the Doppler phase, and $F_{+,k}$ and $F_{\times,k}$ are antenna pattern functions of the $k$-th detector for each polarization mode~\cite{FIM1994}. The covariance of the source parameters is given by the inverse of the Fisher information matrix by
\begin{equation}
  \sqrt{\langle(\delta \lambda^i \delta \lambda^j)\rangle} = \sqrt{(\Gamma^{-1})^{ij}} .
\end{equation}

For our sky localization error calculation, we chose to use source parameters similar to GW170817, as listed in \hyou{SourceParams}, and computed the inspiral waveform to 3.0 and 3.5 post-Newtonian order in amplitude and phase, respectively, compiled in Ref.~\cite{PhenomD}. The Fisher information matrix is computed for eleven parameters, including the nine parameters in \hyou{SourceParams} and two parameters for time and phase at coalescence. The sky localization error is given by
\begin{equation}
 \Delta \Omega_\s = 2 \pi \left| \sin{\theta_\s} \right| \sqrt{(\Delta \theta_\s)^2(\Delta \phi_\s)^2 - \langle \delta \theta_\s \delta \phi_\s \rangle^2},
\end{equation}
where $\Delta \theta_\s \equiv \langle(\delta \theta_\s)^2\rangle^{1/2}$ and $\Delta \phi_\s \equiv \langle(\delta \phi_\s)^2\rangle^{1/2}$. The sky localization error was calculated for 108 uniformly distributed sets of the source location and the polarization angle $\{ \theta_\s, \phi_\s, \psi_{\rm p} \}$, and the median value was used as the objective value to be minimized.

We considered the global network of four gravitational wave detectors: two Advanced LIGO detectors at Hanford and Livingston, Advanced Virgo, and KAGRA. Advanced LIGO detectors are assumed to have its design sensitivity, and Advanced Virgo is assumed to have its binary neutron star optimized sensitivity, as given in Ref.~\cite{ObservationScenario}. We set $f_{\min}=30~\rm{Hz}$ and $f_{\max}$ via \shiki{fISCO} for computing the Fisher information matrix.

\begin{table}
\caption{\label{SourceParams} GW170817-like source parameters assumed for Fisher information matrix analysis. 108 sets of $\{ \theta_\s, \phi_\s, \psi_{\rm p} \}$ are used for sky localization error estimation.}
\begin{ruledtabular}
\begin{tabular}{llcc}
 & Value  \\
\hline
chirp mass & $\chirpmass=1.188 \Msun$ \\
symmetric mass ratio & $\eta=0.248$ \\
luminosity distance & $D_{\rm L} = 40 \unit{Mpc}$ \\
inclination angle & $\iota=28^{\circ}$ \\
colatitude & $\theta_\s$ \\
longitude  & $\phi_\s$ \\
polarization angle & $\psi_{\rm p}$ \\
symmetric spin & $\chi_\s=0^{\circ}$ \\
asymmetric spin & $\chi_{\rm a}=0^{\circ}$ \\
\end{tabular}
\end{ruledtabular}
\end{table}

\section{Particle swarm optimization} \label{SecPSO}
In particle swarm optimization, parameter sets are called particles and their positions $x$ in the multidimensional parameter space are evaluated based on the objective function $\obj(x)$. Their positions are then updated by step sizes called velocities. Their velocities are adjusted based on the distance between the current position and personal best position or swarm's best position, and the iteration stops with a certain termination criterion.

In this section, we describe our implementation and our procedure for tuning the design variables of PSO. There exists many variations of PSO, but here we adopt one of the simplest forms originally proposed by Kennedy and Eberhart in 1995~\cite{PSO1995}.

\subsection{PSO algorithm}
The position of the $k$-th particle at step $(t+1)$ is given by
\begin{equation}
 x_k (t+1) = x_k(t) + v_k(t) ,
\end{equation}
where $v_k(t)$ is its velocity. The velocity is calculated by
\begin{equation}
  v_k(t+1) = w v_k(t) + c_1 r_1 \left(\hat{x}_k - x_k(t) \right) + c_2 r_2 \left( \hat{x}_g - x_k(t) \right) .
\end{equation}
Here $w$ is called the inertia coefficient, and $c_1$ and $c_2$ are called acceleration coefficients. $r_1$ and $r_2$ are two random numbers drawn independently at each step for each particle from uniform distribution in the range [0, 1]. $\hat{x}_k$ is the personal best position which gives the maximum  $\obj(x_k(t))$ over the past positions of the $k$-th particle ({\it pbest$_k$}), and $\hat{x}_g$ is the global best position among all the past positions of the particles ({\it gbest}).

$w$ is usually set slightly smaller than 1, and $c_1$ and $c_2$ are usually set close to 1. Larger $w$ makes the particle move in a straight line, and larger $c_1$ and $c_2$ makes the possibility of the particle overshooting the target positions larger. We use the values suggested in Standard PSO 2006~\cite{StandardPSO2006} as
\begin{equation}
 w = \frac{1}{2 \log{(2)}} = 0.72 ,
\end{equation}
and
\begin{equation}
 c_1=c_2=0.5+\log{(2)}=1.19 .
\end{equation}
We have also tried different values but we did not find any significant improvement in the probability of convergence or computational cost.

\subsection{Initial condition}
We assign uniformly random positions and velocities to particles in our search range $[x_{\min},x_{\max}]$ for the initial step,
\begin{equation}
 x_k(t=0) = x_{\min} + r (x_{\max} - x_{\min})
\end{equation}
and
\begin{equation}
 v_k(t=0) = (r-0.5) (x_{\max} - x_{\min}) ,
\end{equation}
where $r$ is a random number drawn independently from uniform distribution in the range [0, 1] for each particle.

\subsection{Boundary condition}
To ensure that the particles search for the global maximum inside the predefined search space, boundary violating particles must be treated appropriately. There have been a number of boundary conditions proposed, and a good summary is provided in Ref.~\cite{PSOboundary}. We use one of the most conventional boundary conditions, the {\it reflecting wall} condition. If a particle crosses a boundary in one of the dimensions, it is relocated at the boundary of that dimension,
\begin{eqnarray}
  \left\{
    \begin{array}{ll}
  x_k^i(t) = x_{\min}^i &{\rm if}~~  x_k^i(t) < x_{\min}^i \\
  x_k^i(t) = x_{\max}^i &{\rm if}~~  x_k^i(t) > x_{\max}^i ,
    \end{array}
  \right.
\end{eqnarray}
and the velocity is reversed for that dimension,
\begin{equation}
  v_k^i(t) = -v_k^i(t) ~~{\rm if}~~ x_k^i(t) < x_{\min}^i ~~{\rm or}~~  x_k^i(t) > x_{\max}^i.
\end{equation}
Here, superscript $i$ indicate the index of the dimension.

\subsection{Termination criterion}
To terminate the computation, we used a simple criterion based on the accuracy we need for optimization. We stop iterating if the change in the global best value $\obj(\hat{x}_g)$ is less than a certain threshold. Depending on the objective function to use, we set the threshold to be $\delta \obj = 10^{-3} \unit{Mpc}$ or $10^{-5} \unit{deg^2}$, which is small enough compared with the precision that is experimentally realizable.

We note here that this does not mean that the resulting objective values of PSO runs always converge within this threshold. Since PSO is a stochastic method, convergence to the true global maximum can only be quantified in terms of probability.

\subsection{Tuning the number of particles}
We are now left with only one PSO variable to be tuned: the number of particles $\Np$. Tuning of $\Np$ was done systematically by following a procedure similar to Ref.~\cite{PSOMohanty2010}, based on the probability of convergence. Unlike the gravitational wave data analysis dealt in Ref.~\cite{PSOMohanty2010}, we are focused more on the objective function values rather than the optimized detector parameters. We therefore calculate the probability of convergence in terms of the resulting objective values. We ran independent PSO runs multiple times to see if the resulting objective values converge within $100 \times \delta \obj = 0.1 \unit{Mpc}$ or $10^{-3} \unit{deg^2}$. The probability of convergence is defined by the fraction of runs in which the resulting objective value is consistent with the best value among the runs within this threshold.

\begin{table}
\caption{\label{ParticleNumber} The mean number of iterations and probability of convergence from 100 independent PSO runs for different number of search parameters $\Nd$ and objective functions.}
\begin{ruledtabular}
\begin{tabular}{lccc}
number of search parameters $\Nd$ & 3   & 5   & 7   \\
number of particles        $\Np$ & 10  & 20  & 200 \\
\hline
\multicolumn{4}{l}{inspiral range optimization} \\
\quad number of iterations       & $52\pm13$ & $73\pm16$ & $60\pm18$ \\
\quad probability of convergence & 98\%      & 96\%      & 91\%      \\
\multicolumn{4}{l}{sky localization optimization} \\
\quad number of iterations       & $28\pm10$ & $47\pm14$ & $38\pm10$ \\
\quad probability of convergence & 99\%      & 92\%      & 98\% \\
\end{tabular}
\end{ruledtabular}
\end{table}

For different number of search parameters $\Nd$, we increased $\Np$ until the probability of convergence reached more than $90\%$, and settled on $\Np=10,~20$, and 200 for $\Nd=3,~5$, and 7, respectively.
\hyou{ParticleNumber} summarizes our result of 100 independent PSO runs for each combination of $\Nd$ and two objective functions. It is reasonable that optimization with $\Nd=7$ requires more $\Np$ than the other two cases since the resonant peaks of the suspension thermal noise changes with $l_\f$ and $d_\f$. To save the computational cost while tuning $\Nd$, the source location and the polarization angle were fixed for the sky localization optimization, instead of calculating the sky localization error for all 108 sets of the angular parameters.

The computational cost can evaluated by the number of objective function evaluations, which equals to $\Np$ times the number of iterations. It is worth mentioning here that as shown in \hyou{ParticleNumber}, the computational cost does not grow exponentially with $\Nd$. Simple grid-based search requires $O(10^5)$, $O(10^9)$, and $O(10^{14})$ objective function evaluations for $\Nd=3,~5$, and 7, respectively, if we want to optimize the detector parameters within 0.1~Mpc.

\begin{table}
\caption{\label{PSODesign} PSO design variables used in this work. The number of particles is tuned based on the probability of convergence and it differs by the number of search parameters.}
\begin{ruledtabular}
\begin{tabular}{ll}
 & Value \\
\hline
inertia coefficient                & $w=0.72$ \\
acceleration coefficients           & $c_1=c_2=1.19$ \\
termination threshold               & $\delta \obj = 10^{-3}\unit{Mpc}$ or $10^{-5} \unit{deg^2}$ \\
number of particles                 & $\Np =10,~20,~200$ \\
number of search parameters         & $\Nd = 3,~5,~7$ \\
\end{tabular}
\end{ruledtabular}
\end{table}

A final set of PSO design variables of our implementation is given in \hyou{PSODesign}.

\begin{table*}
\caption{\label{ResultParams} Optimized KAGRA parameter values and obtained objective function values for both inspiral range optimization and sky localization optimization, with different number of search parameters $\Nd$. Inspiral range optimization with $\Nd=3$ corresponds to current KAGRA default design sensitivity (see \hyou{SearchRange}). The parameter values in the parenthesis indicate that they are fixed parameters not used for optimization. Input power at BS, which is a function of $T_\m,~I_{\attn},~l_\f$, and $d_\f$ is also shown.}
\begin{ruledtabular}
\begin{tabular}{llcccccc}
 & & \multicolumn{3}{l}{Inspiral range optimization} & \multicolumn{3}{l}{Sky localization optimization} \\
 & & $\Nd=3$ & $\Nd=5$ & $\Nd=7$                     & $\Nd=3$ & $\Nd=5$ & $\Nd=7$ \\
\hline
detuning angle (deg)   & $\phidet$               & 3.5   & 3.5   & 3.5   & 0.4   & 1.1   & 1.0   \\
homodyne angle (deg)   & $\zeta$                 & 134.4 & 114.9 & 116.0 & 100.7 & 117.0 & 119.8 \\
mirror temperature (K) & $T_\m$                  & 21.6  & 22.9  & 20.5  & 30.0  & 30.0  & 26.9  \\
power attenuation      & $I_{\attn}$             & (1)   & 1.0   & 1.0   & (1)   & 1.0   & 1.0   \\
SRM reflectivity (\%)  & $R_{\SRM}$              & (84.6)& 93.8  & 96.5  & (84.6)& 93.1  & 96.4  \\
fiber length (cm)      & $l_\f$                  & (35)  & (35)  & 24.0  & (35)  & (35)  & 20.0  \\
fiber diameter (mm)    & $d_\f$                  & (1.6) & (1.6) & 2.2   & (1.6) & (1.6) & 2.5   \\
input power at BS (W)  & $I_0$                   & 616   & 834   & 1760  & 2600  & 2600  & 11200 \\
\hline
\multicolumn{2}{l}{$1.4\Msun$-$1.4\Msun$ inspiral range (Mpc)}      & {\bf 152.8} & {\bf 158.1} & {\bf 168.7} & 124.7 & 134.8 & 149.4 \\
\multicolumn{2}{l}{median sky localization error (${\rm deg^2}$)}   & 0.186 & 0.186 & 0.167 & {\bf 0.142} & {\bf 0.137} & {\bf 0.107} \\
\end{tabular}
\end{ruledtabular}
\end{table*}

\section{Optimization results} \label{SecResults}

The results of KAGRA sensitivity optimization for binary neutron star inspiral range and sky localization of GW170817-like binary are summarized in \hyou{ResultParams}, and optimized sensitivity curves are shown in \zu{ResultSensitivity}.

\subsection{Inspiral range optimization}
We can see that the result of inspiral range optimization with $\Nd=3$ is consistent with the KAGRA default values in \hyou{SearchRange}. The default values are determined by grid-based search, and PSO successfully gave consistent results within the accuracy of the grid size, which was $0.1^{\circ}$, $0.1^{\circ}$, and $1\unit{K}$ for $\phidet$, $\zeta$, and $T_\m$, respectively.

The result with $\Nd=5$ show that the inspiral range can be improved by 4\%, by replacing the SRM to one with a higher reflectivity of 93.8\%. This was also pointed out in Ref.~\cite{AsoKAGRA}, but we have chosen to use $R_{\SRM}=85\%$ as the default, since a SRM with higher reflectivity gives worse inspiral range in the tuned SRC ($\phidet=0$) case.

The optimization result with all the seven parameters show that the inspiral range can be improved by 10\% by simply changing the parameter values. This means a roughly 30\% improvement in the detection rate, since the detection rate is proportional to the cubic of the inspiral range. This improvement is given by changing the test mass suspension fibers to shorter and thicker ones to increase the input power, while keeping the mirror temperature low. This is effective for reducing both thermal noise and shot noise in the mid-frequencies of the observation band.

We also see that inspiral range optimization results in high SRC detuning. This gives a narrower observation band and better sensitivity at mid-frequencies. This is optimal for improving the signal to noise ratio and increasing the inspiral range, but this is not optimal for sky localization, as discussed in the next subsection.

\subsection{Sky localization optimization}
As apparent from \hyou{ResultParams}, the sky localization optimization generally result in almost no detuning of SRC, a homodyne phase close to conventional phase quadrature readout, and higher test mass temperature. This can be understood by considering that the frequency of the gravitational waves at the innermost stable circular orbit of GW170817 is $1.6\unit{kHz}$. For sky localization, coalescence timing measurement between the detectors around the globe is important. Therefore, broadband detection and reducing the shot noise at higher frequencies by increasing the input power at the cost of thermal noise increasing at lower frequencies is effective for sky localization of binary neutron stars.

The optimization result with all the seven parameters show that the median value of the sky localization error can be reduced to $0.1 \unit{deg^2}$, from the default $0.2 \unit{deg^2}$. The sky localization improvement for the uniformly distributed 108 sets of the source location and the polarization angle was a factor of $1.6 \pm 0.2$ on average.

This is possible by making the test mass suspension fibers as short and thick as possible within the search range, while keeping the mirror temperature low enough to reduce the mirror thermal noise. These changes allow the input power at the BS to increase to 11~kW. Since KAGRA's power recycling gain is 10, this requires a laser source with power at $\sim 1.1 \unit{kW}$. Currently, this is not technically feasible, but the input power can be effectively increased by squeezed vacuum injection~\cite{LIGOSqueezing}. Squeezed vacuum injection also relaxes the requirement for heat extraction of the test masses, and therefore helps in reducing the test mass temperature. We leave incorporation of the squeezing parameters for optimization, as well as more detailed experimental feasibility study for our future work.

We also point out that this high frequency shift of the observation band results in the degradation in the inspiral range. This reduces the detection rate of KAGRA alone by 50\% for $\Nd=3$ case. However, the detection rate by the global network would not be reduced as much. Therefore, KAGRA focusing on high frequencies might be an option in the global network era.

For both inspiral range optimization and sky localization optimization, it is shown that the input power should be as high as possible ($I_{\attn}=1$) for our search range. This is because reducing shot noise at higher frequencies is critical for increasing the signal to noise ratio for binary neutron stars. The result would change if sensitivity optimization is done for binary black holes, which merge at lower frequencies.

\def\miniwid{0.9\hsize}
\begin{figure*}
	\begin{center}
\begin{minipage}[b]{0.49\hsize}
   \begin{center}
   \includegraphics[width=\miniwid]{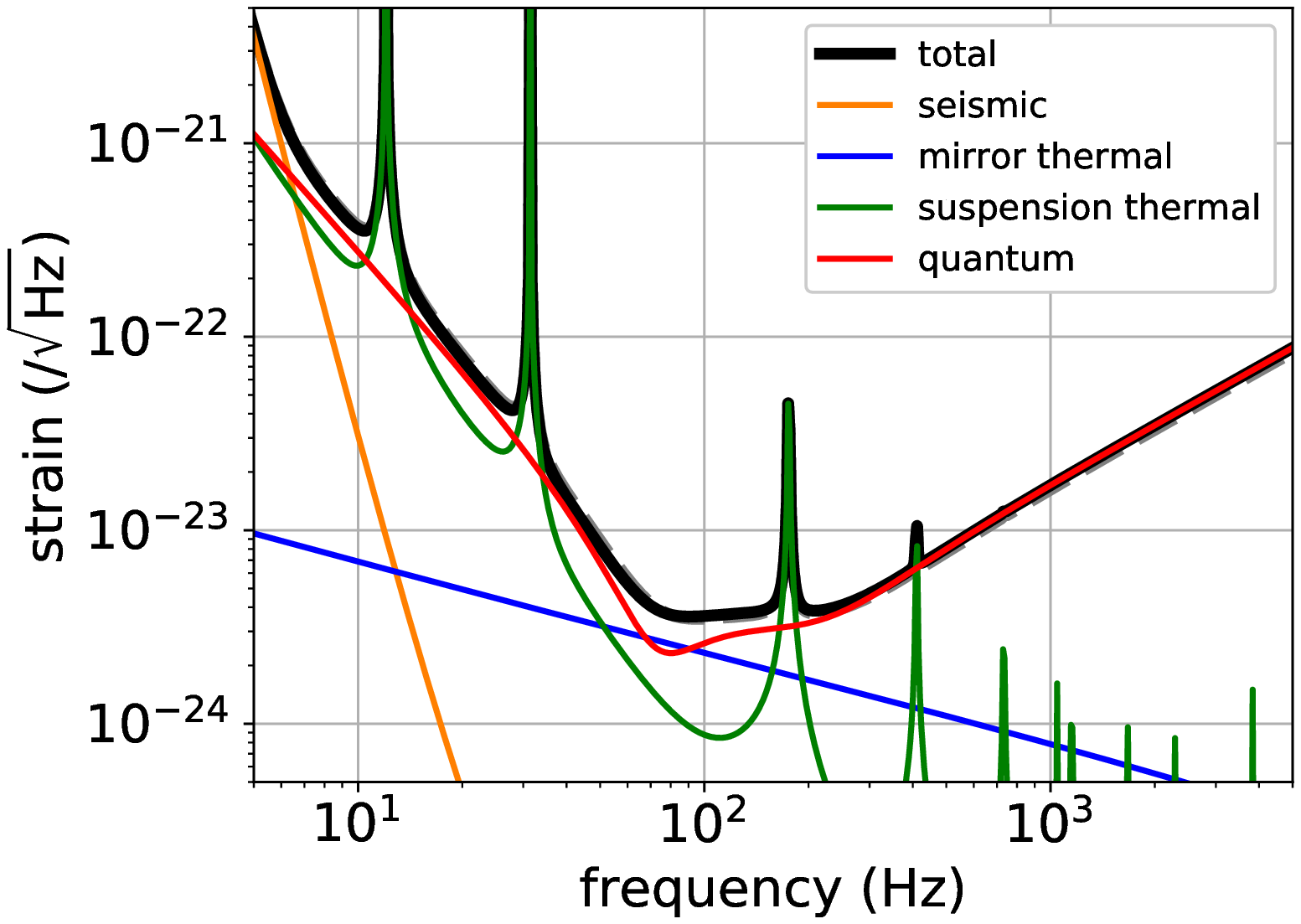} \\
   (a) Inspiral range optimization, $\Nd=3$ \\
   \includegraphics[width=\miniwid]{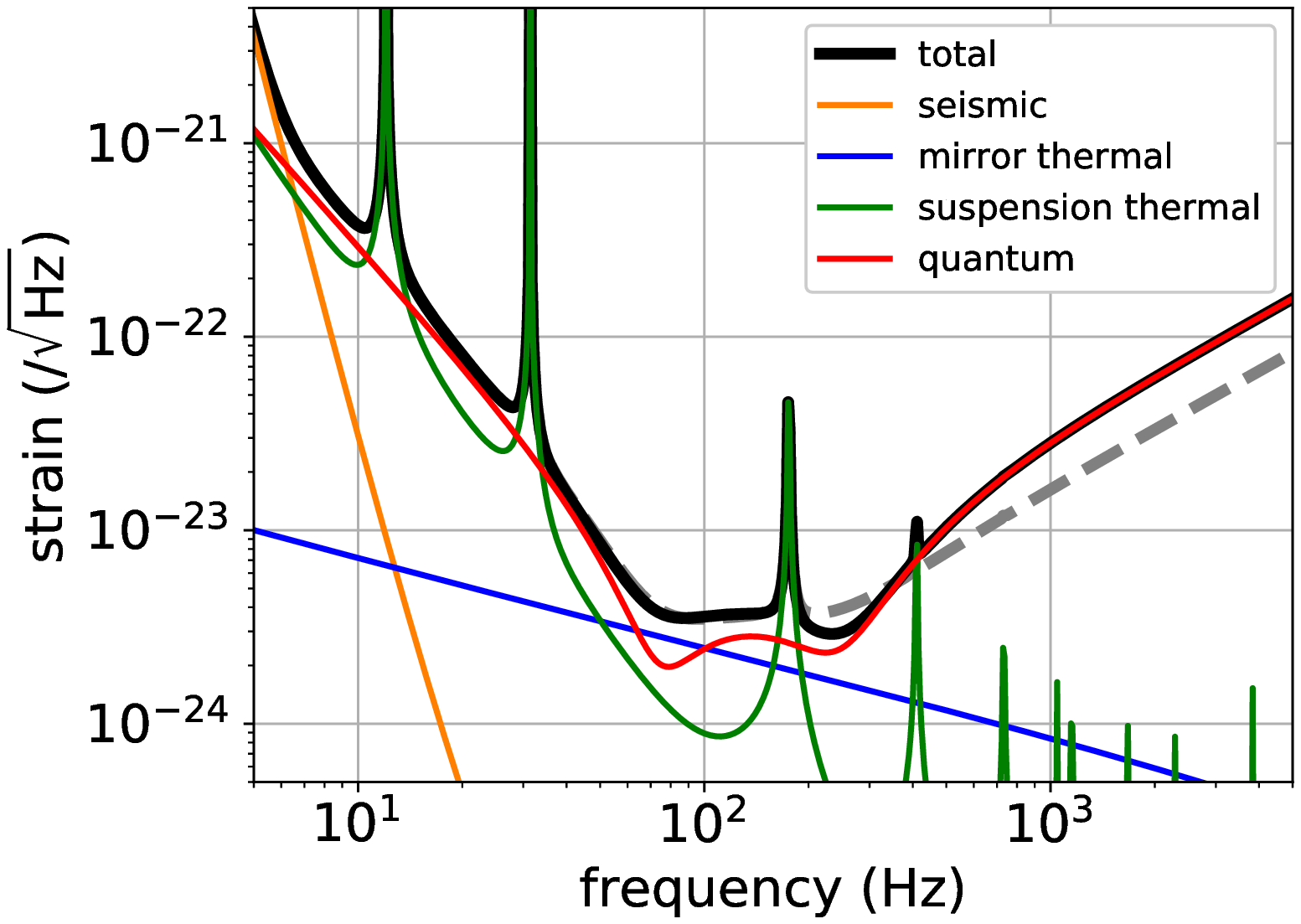} \\
   (b) Inspiral range optimization, $\Nd=5$ \\
   \includegraphics[width=\miniwid]{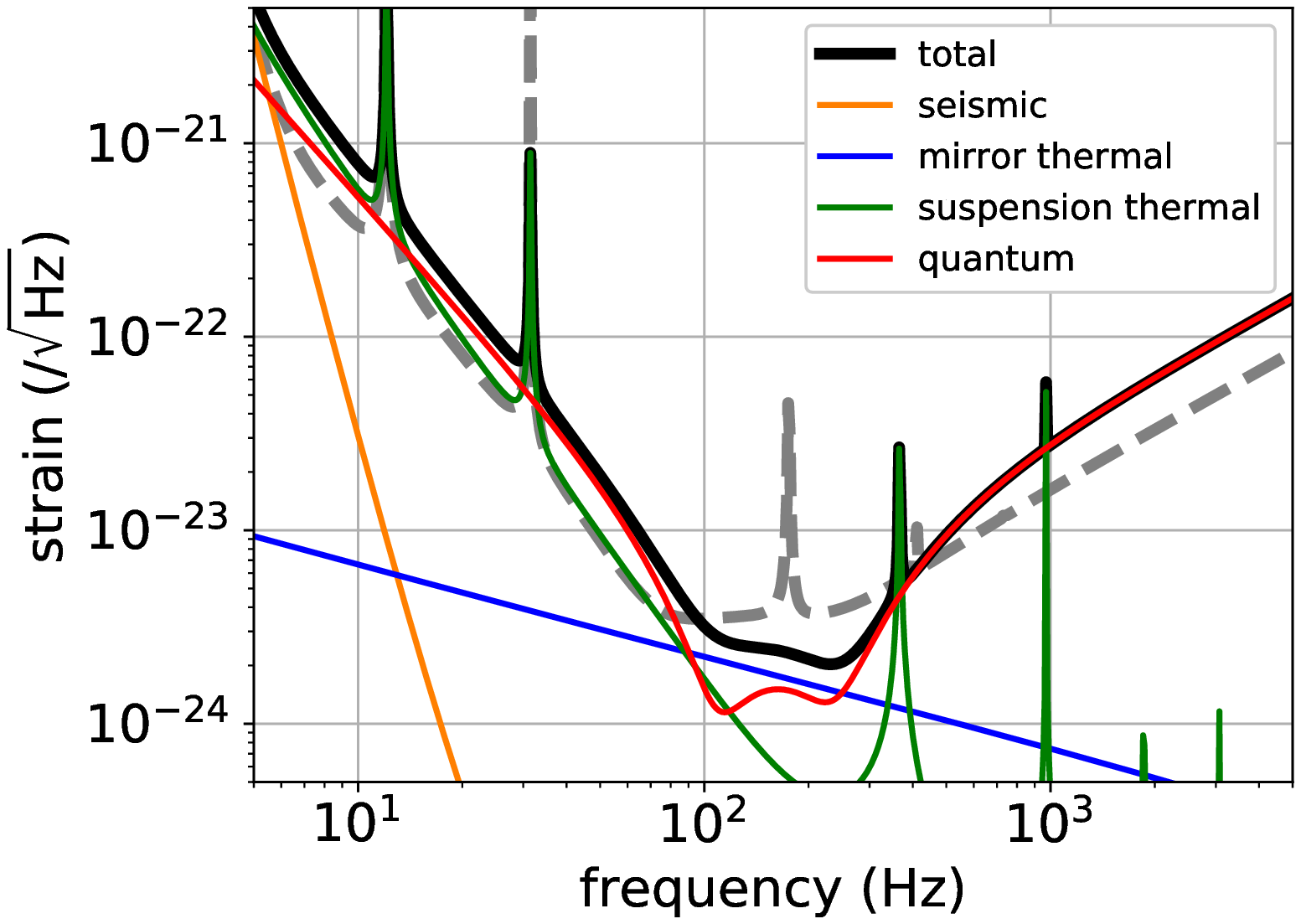} \\
   (c) Inspiral range optimization, $\Nd=7$
   \end{center}
\end{minipage}   
\begin{minipage}[b]{0.49\hsize}
   \begin{center}
   \includegraphics[width=\miniwid]{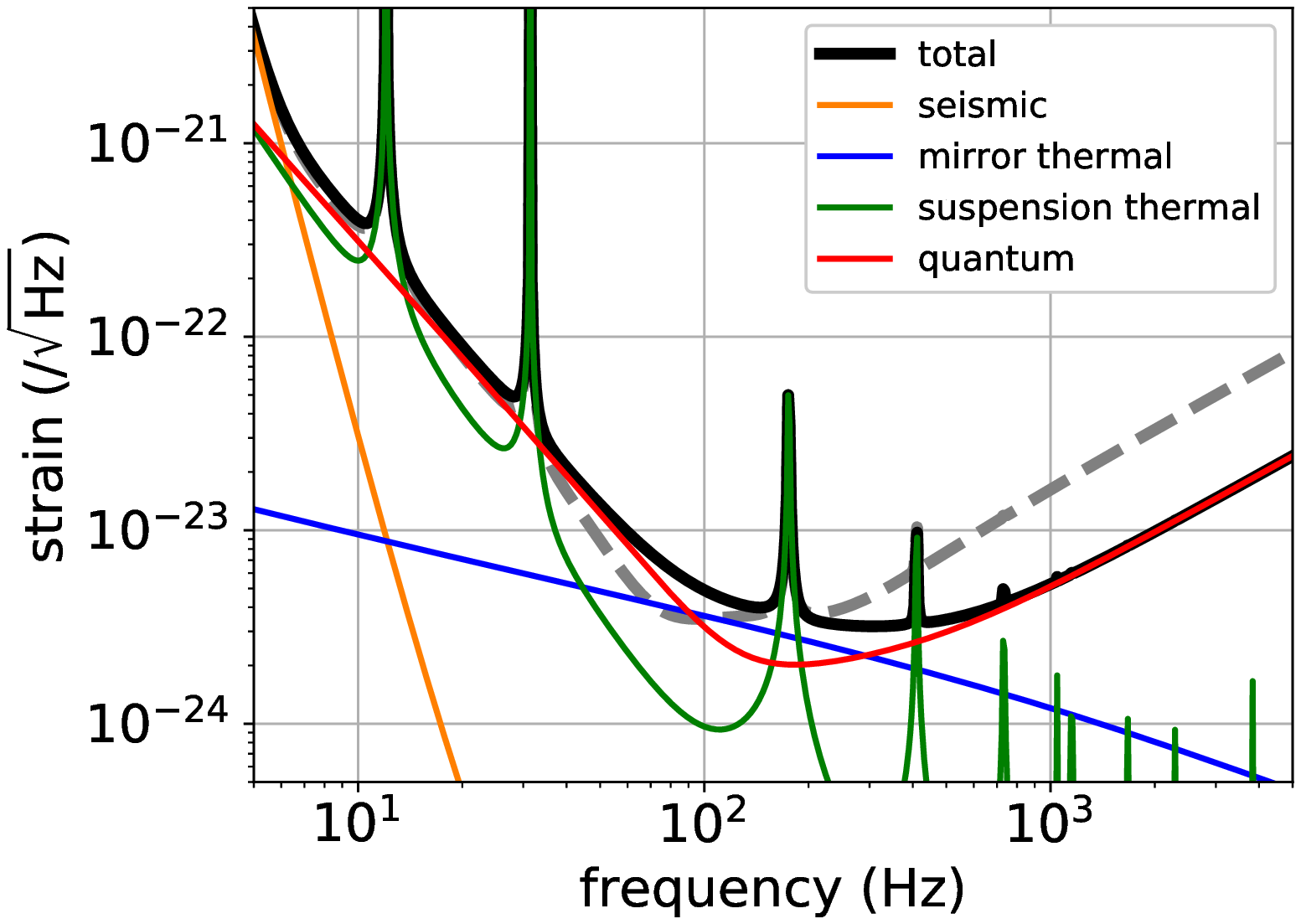} \\
   (d) Sky localization optimization, $\Nd=3$ \\
   \includegraphics[width=\miniwid]{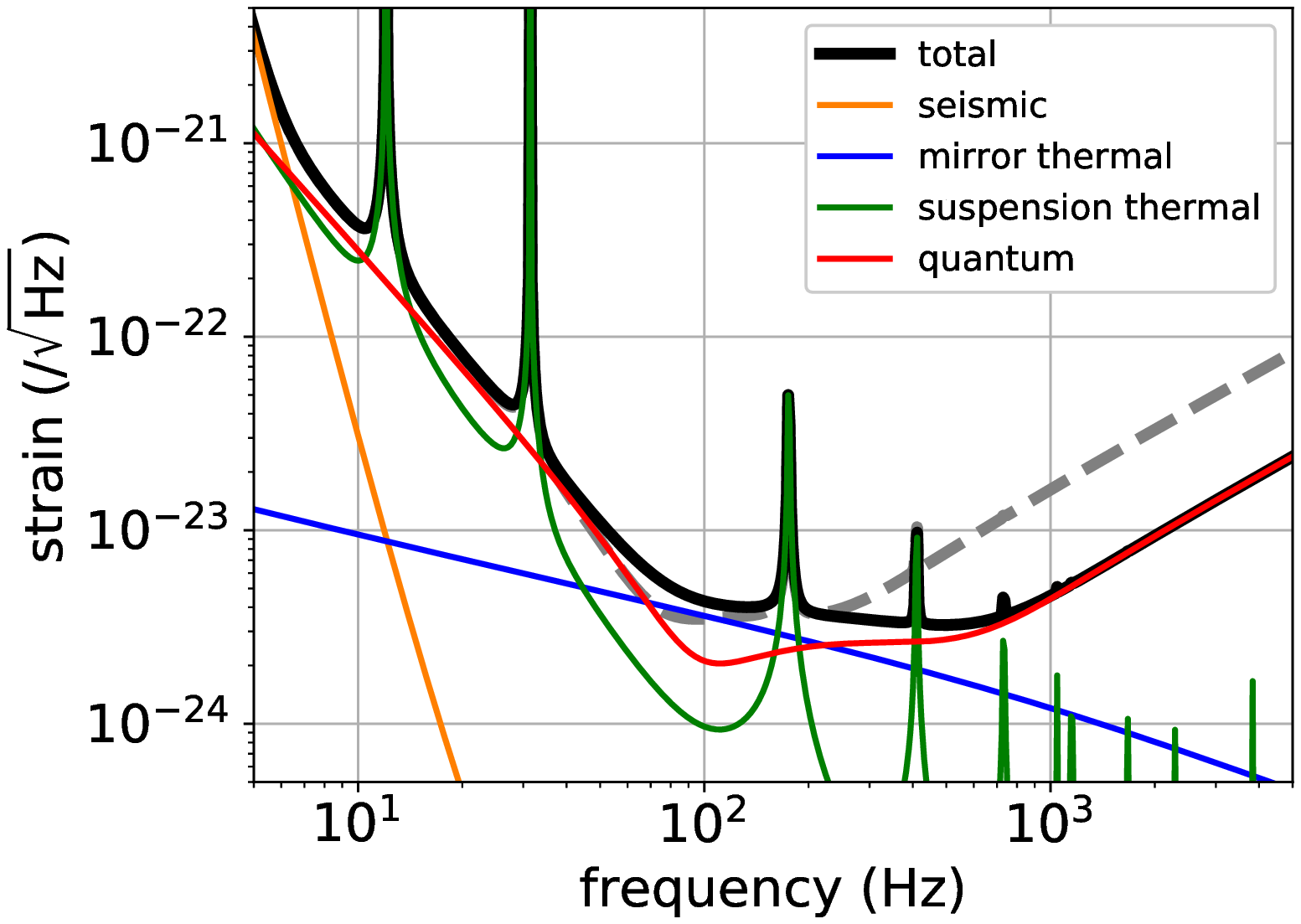} \\
   (e) Sky localization optimization, $\Nd=5$ \\
   \includegraphics[width=\miniwid]{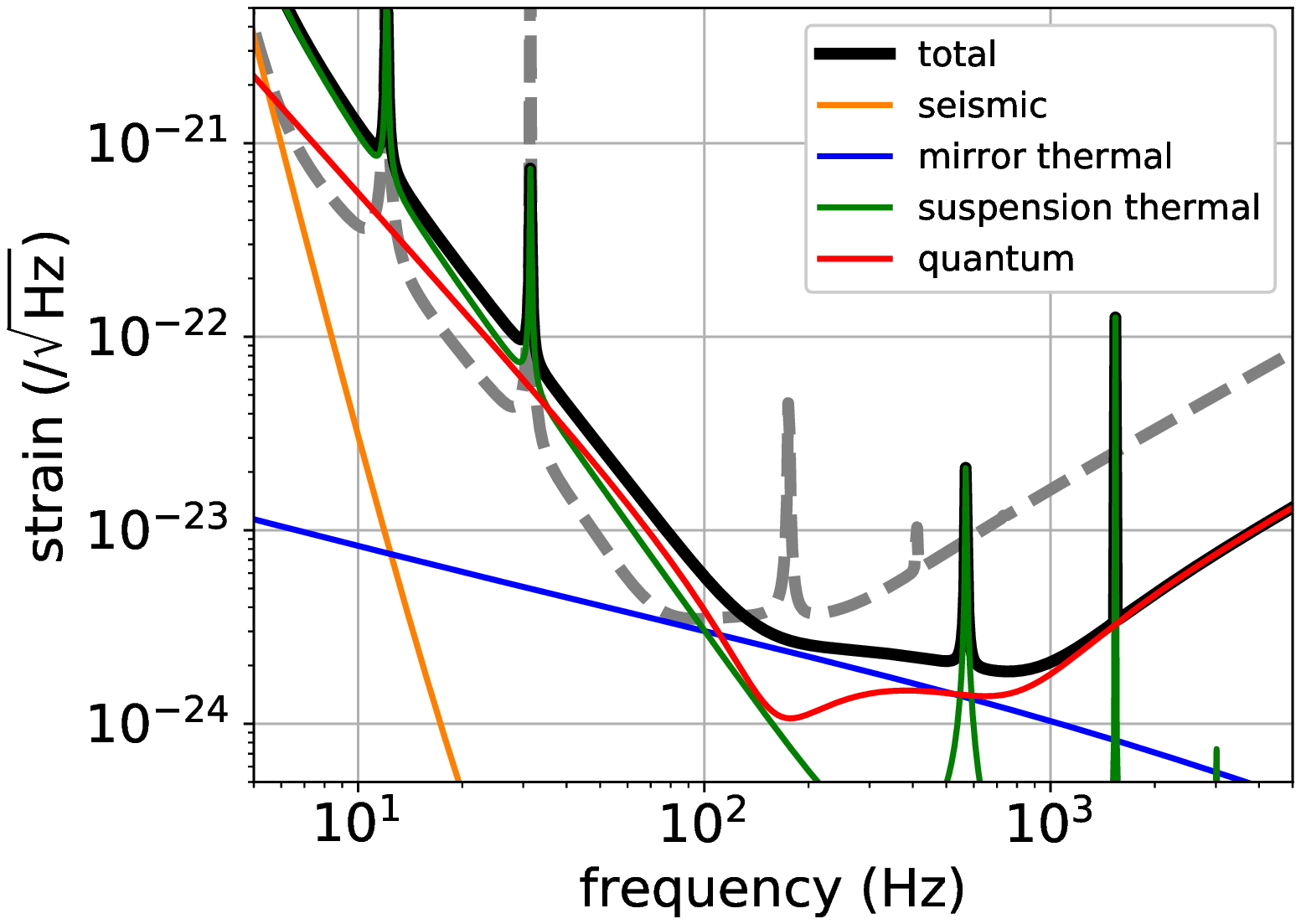} \\
   (f) Sky localization optimization, $\Nd=7$
   \end{center}
\end{minipage}
	\caption{\label{ResultSensitivity} Optimized sensitivity curves for both inspiral range optimization and sky localization optimization, with different number of search parameters $\Nd$. For comparison, the KAGRA default sensitivity calculated with default parameters in \hyou{SearchRange} is plotted with gray dashed line.}
	\end{center}
\end{figure*}

\section{Conclusion} \label{SecConclusion}
We performed the first application of PSO to the sensitivity design of a cryogenic gravitational wave detector. Our results from PSO successfully showed that binary neutron star inspiral range and sky localization of the KAGRA detector can be improved just by retuning the parameters of already existing components. The improvement for optimization using seven parameters was $10\%$ for the binary neutron star inspiral range, and a factor of 1.6 for sky localization of GW170817-like binary averaged across the sky. By running PSO with different number of search parameters, we also confirmed that the computational cost does not grow with number of dimensionality of the parameter space.

It is expected that future gravitational wave detectors will have more detector parameters, which need to be optimized. It is also expected that figures of merit other than the inspiral range will be important in the era of gravitational wave astronomy. PSO is a generic optimization method and can be applied to a variety of objective functions. Our results show that PSO is effective for the sophisticated design of future gravitational wave detectors.

\section*{Acknowledgements}
We thank Takahiro Yamamoto, Sadakazu Haino, and Kazuhiro Yamamoto for independently verifying the codes used in this paper.
We would also like to thank Nobuyuki Matsumoto and Ooi Ching Pin for fruitful discussions.
K. K., H. T. and Y. E. acknowledge financial support received from the Advanced Leading Graduate Course for Photon Science (ALPS) program at the University of Tokyo.
This work was supported by JSPS Grant-in-Aid for Young Scientists (A) No. 15H05445.

The KAGRA project is supported by MEXT, JSPS Leading-edge Research Infrastructure Program, JSPS Grant-in-Aid for Specially Promoted Research 26000005, MEXT Grant-in-Aid for Scientific Research on Innovative Areas 24103005, JSPS Core-to-Core Program, A. Advanced Research Networks, the joint research program of the Institute for Cosmic Ray Research, University of Tokyo, National Research Foundation (NRF) and Computing Infrastructure Project of KISTI-GSDC in Korea, the LIGO project, and the Virgo project.

%

\end{document}